\documentclass[11pt,a4paper]{amsart}

\usepackage[latin1]{inputenc}
\usepackage[english]{babel}
\usepackage{amsmath}
\usepackage{bm}
\usepackage{amssymb}
\usepackage{graphicx}
\usepackage{braket}
\usepackage[pdftex,plainpages=false,colorlinks,hyperindex,bookmarksopen,linkcolor=red,citecolor=blue,urlcolor=blue]{hyperref}
\usepackage{cite}
\usepackage{mathrsfs}
\usepackage{epstopdf}
\usepackage{braket}
\usepackage{cleveref}
\usepackage[hmargin=3cm,vmargin={3.5cm,4cm}]{geometry}

\theoremstyle{theorem}

\theoremstyle{definition}                           

\theoremstyle{remark}                             
\newtheorem{remark}{Remark}[section]              

\usepackage{color}

\usepackage{mathtools,slashed}

\newcommand{\be}{\begin{eqnarray}}
\newcommand{\ee}{\end{eqnarray}}
\def\eg{{\em e.g.}}
\def\ie{{\em i.e.}}
\def\Rset{{\mathbb{R}}}

\def\eu{{\ensuremath{\mathrm{e}}}}

\renewcommand{\Re}{\operatorname{Re}}

\renewcommand{\d}{\mbox{${\rm d}$}} 

\newcommand{\wh}[1]{\widehat{#1}}

\newcommand{\Gn}{G_{\rm N}}

\allowdisplaybreaks
\begin{document}

\title[On the Kuzmin model in fractional Newtonian gravity \dots]{On the Kuzmin model in \\ fractional Newtonian gravity}
	
		\author[A. Giusti]{Andrea Giusti${}^{1}$}
		\address{${}^{1}$ 
		Bishop's University,
		Physics $\&$ Astronomy Department, 
		2600 College Street, Sherbrooke, J1M 1Z7,
		QC	CANADA}	
 		\email{agiusti@ubishops.ca}
		
		\author[R. Garrappa]{Roberto Garrappa${}^{2}$}
		\address{${}^{2}$ 
		Department of Mathematics,
 		University of Bari, 
		Via E. Orabona 4, 70126 Bari, ITALY
		and 
		the INdAM Research group GNCS}	
 		\email{roberto.garrappa@uniba.it}

 		\author[G. Vachon]{Genevi{\`e}ve Vachon${}^{3}$}
		\address{${}^{3}$ Bishop's University,
		Physics $\&$ Astronomy Department, 
		2600 College Street, Sherbrooke, J1M 1Z7,
		QC	CANADA}	
 		\email{gvachon18@ubishops.ca}
	
\begin{abstract}
Fractional Newtonian gravity, based on the fractional generalization of Poisson's equation for Newtonian gravity, is a novel approach to Galactic dynamics aimed at providing an alternative to the dark matter paradigm through a non-local modification of Newton's theory. We provide an in-depth discussion of the gravitational potential for the Kuzmin disk within this new approach. Specifically, we derive an integral and a series representation for the potential, we verify its asymptotic behavior at large scales, and we provide illuminating plots of the resulting equipotential surfaces.
\end{abstract}

\thanks{Published in: {\em Eur. Phys. J. Plus} \textbf{135} (2020) 798; 
doi: \href{https://doi.org/10.1140/epjp/s13360-020-00831-9}{10.1140/epjp/s13360-020-00831-9}}

\maketitle

%
%
%
%

\section{Introduction}
\label{sec:into}
	
	Galaxy rotation curves and the formation of large-scale structure in the universe are among the most compelling indications that General Relativity and the Standard Model of particle physics cannot account for all natural phenomena. The situation is even more severe than that, indeed it turns out that the theoretical tools which are currently available in physics can only resolve about 5\% of the content of the universe. In more detail, in order to explain the current accelerating expansion of the universe it is customary to postulate the existence of an exotic {\em dark energy} \cite{Amendola, Brax:2017idh} fluid, with positive energy and negative pressure, affecting the universe on its largest scale. Similarly,  in order to account for structure formation after the Big Bang, as well as deviations from the expected Newtonian predictions for galaxy rotation curves, it seems to be necessary to include an additional dark component of the universe, featuring no direct coupling with electromagnetic radiation and an (almost) imperceptible pressure, which is dubbed as {\em dark matter} \cite{Bertone:2004pz, Garrett:2010hd, Bertone:2016nfn}. In the picture discussed above dark matter and dark energy are treated as exotic forms of matter evading the Standard Model of particle physics. This exotic matter content finds its way in the so-called standard model of cosmology, also know as the $\Lambda$--Cold Dark Matter model or $\Lambda$CDM for short, according to which the energy content of the universe splits into a 5\% of ordinary (luminous) matter, 25\% of dark matter, and about 70\% is accounted for by dark energy. It is worth stressing that cold dark matter, namely dark matter moving with non-relativistic velocity, seems to be favored with respect to ``warm'' and ``hot'' models since it yields predictions for the cosmological large-scale structure that generally agree with current astronomical observations \cite{Blumenthal:1984bp}.

	An alternative approach to dark matter and dark energy, which are typically added {\em ad hoc} in Einstein's theory to reproduce the astronomical and cosmological observations, requires to rethink gravitational physics at a more fundamental level and include large-scale modifications of gravity aimed at reconciling theory and experiments. Notably, extensive efforts have been devoted toward the study of alternative theories of gravity that could replace, at least in part, dark matter and dark energy with the phenomenology of some additional gravitational degrees of freedom, see {\eg}, \cite{Capozziello:2003tk, Carroll:2003wy, 
Sotiriou:2008rp, DeFelice:2010aj, Nojiri:2010wj, Capozziello:2011et, Capozziello:2009nq, Nojiri:2017ncd, Nojiri:2008nt, Nojiri:2009kx, Heisenberg:2018vsk}. However, lacking a direct detection of new particles signaling the emergence of physics beyond the Standard Model, and any definitive experimental proof of significant deviations from General Relativity, one can only conclude the jury is still out on what is really responsible for the odd phenomena observed at galactic and cosmological scales.

	One of the most successful proposal of modification of gravity theory aimed at explaining the phenomenology typically traced back to dark matter is known as Modified Newtonian Dynamics (MOND), originally introduced by M. Milgrom in \cite{Milgrom:1983ca, Milgrom:1983pn, Milgrom:1983zz, Bekenstein:1984tv}. The idea behind this approach relies on the assumption that there exists a critical acceleration scale $a_0$, whose value is empirically determined, such that Newton's gravity dramatically changes when the magnitude of the acceleration of a test particle falls below this threshold. Specifically, under the simple assumption of spherical symmetry and considering a test particle on a stable orbit around a core mass $M$, denoting by $a = a(r)$ the acceleration of the test body MOND predicts that for $a \gg a_0$ one recovers standard Newtonian gravity, {\ie}, 
\be
a \simeq \frac{\Gn \, M}{r^2} \, ,
\ee
whereas when $a \ll a_0$ the dynamics of the test particle is modified according to
\be
\frac{a^2}{a_0} \simeq \frac{\Gn \, M}{r^2} \, 
\ee
with $r$ denoting the distance from the center of the system. In other words, MOND recovers the standard Newtonian scaling of the acceleration $a(r) \sim 1/r^2$ at short scales, whilst the model yields the asymptotic behavior $a(r) \sim 1/r$ at large (Galactic) scales. This implies that the rotational velocity of a test body around a Galaxy center behaves as $v^2(r) \sim \Gn \, m(r)/r$ in the innermost part of the Galaxy, with $m(r)$ denoting the total mass contained within a circular orbit of radius $r$, while $v^4(r) \sim \Gn \, M \, a_0$ as one moves away from the Galaxy center. On other words, galaxy rotation curves flatten out as one moves asymptotically far from the Galaxy center, in full agreement with various astronomical observations \cite{Garrett:2010hd, Begeman:1989kf, Zwicky:1937zza, Corbelli:1999af}. In \cite{Bekenstein:1984tv} J. D. Bekenstein and M. Milgrom proposed a non-relativistic potential theory reproducing the MOND scenario based on a non-linear modification of the Poisson equation of Newtonian gravity. The first robust relativistic MOND inspired model, known as Tensor--Vector--Scalar gravity or TeVeS, was then proposed by J. D. Bekenstein in \cite{Bekenstein:2004ne}. Clearly, this last proposal is not exempt from problems, however in the broader scheme of things it served as the seminal work for the study of dark matter phenomenology as an emergent effect of alternative theories of gravity.

	Fractional calculus \cite{FC1, FC2, FC3} offers a reliable set of tools for describing several physical phenomena which are not typically accounted for by model based on ordinary calculus (see \eg,\cite{FC3, FC4, FC5}). In recent years, this mathematical scheme has also been applied, in various forms, to gravity and fundamental physics, see \eg, \cite{Calcagni:2011kn, Tarasov:2018zjg, Barvinsky:2019spa, Frassino:2019yip}. Focusing on the problem of dark matter phenomenology, the first {\em fractional} MOND--like non-relativistic potential theory was proposed by A. Giusti in \cite{mine}. This approach is based on a fractional modification of the Poisson equation of Newtonian gravity, where the ordinary Laplacian $-\triangle$ is replaced by the so called fractional Laplacian $(-\triangle) ^s$ with $s \in [1, 3/2)$. Notably, another model for a MOND--like non-relativistic potential theory, somehow related to fractional calculus, was proposed by G.~U.~Varieschi in \cite{Varieschi:2020ioh, Varieschi:2020dnd}. Varieschi's approach is very similar to the one in \cite{mine} thought the two are not identical as discussed in \cite{Varieschi:2020ioh}. The key difference lays in the fact that Varieschi's model is {\em not a fractional theory}. Indeed, Varieschi's model relies on the use of a generalized gravitational Gauss's law where the standard integration over $\Rset^3$ is replaced with an Hausdorff measure of $\Rset^3$ related to Weyl's fractional integral. This procedure turns the model into a generalization of Newtonian gravity on a {\em fractal} space, involving a measure inspired by fractional calculus, for which {\em the field equation remains of integer order} (and thus {\em local}). This specific caveat, however, does not make Varieschi's model any less interesting or less deserving of further investigation.

	This work is organized as follows: first, we recall the basics of Giusti's {\em fractional Newtonian gravity}, introduced in \cite{mine}, and its implications for Galactic dynamics; second, we review the preliminary results discussed in \cite{mine} for the Kuzmin disk; third, we complete the analysis for the Kuzmin disk in fractional Newtonian gravity providing a discussion of the asymptotic behavior of the corresponding potential, a series representation for the full potential outside the plane of the disk, and a numerical study of the equipotential surfaces as one varies the fractional parameter $s \in [1, 3/2)$.

%
%
%
%

\section{Fractional Newtonian Gravity}
\label{sec:FNG}

Fractional Newtonian gravity \cite{mine} is an alternative to standard Newtonian gravity based on a modification of the Poisson equation for the gravitational potential. Specifically, the key ingredient of this model consist in the so-called {\em fractional Laplacian}. 

	Let $f(\bm{x})$ be a sufficiently well-behaved function on $\Rset ^3$, one defines the Fourier transform of $f(\bm{x})$ as
\be
\wh{f} (\bm{k}) \equiv \mathcal{F} \left[ f (\bm{x}) \, ; \, \bm{k} \right] = 
\int _{\Rset ^3} \eu ^{- i \bm{k} \cdot \bm{x}} \, f (\bm{x}) \, \d ^3 x \, ,
\ee
with $\cdot$ denoting the standard Euclidean scalar product on $\Rset ^3$. Hence, if 
$$\triangle f (\bm{x}) := {\rm div}[\bm{\nabla} f (\bm{x})]$$
denotes the Laplacian of $f(\bm{x})$, then it is easy to see that
\be
\mathcal{F} \left[(-\triangle) f (\bm{x}) \,  ; \, \bm{k} \right] = |\bm{k}|^2 \, \wh{f} (\bm{k}) \, ,
\ee
with $|\bm{k}|^2 \equiv \bm{k} \cdot \bm{k}$. The {\em fractional Laplacian} \cite{FL1, FL2} is therefore defined as the operator $(-\triangle) ^s$ such that
\be
\label{eq:FourierFL}
\mathcal{F} \left[(-\triangle)^s f (\bm{x}) \,  ; \, \bm{k} \right] = |\bm{k}|^{2s} \, \wh{f} (\bm{k}) \, .
\ee
Ten equivalent representations of this operator are discussed in \cite{FL2}. Further details on the fractional Laplacian are analyzed and reviewed in \cite{Karniadakis}.

Fractional Newtonian gravity \cite{mine} is therefore based on the fractional Poisson equation
\be
\label{eq:fractionalpoisson}
(- \triangle) ^s \Phi (\bm{x}) = - 4 \, \pi \, \Gn \, \ell ^{2 - 2 s} \, \rho (\bm{x}) \, ,  
\ee
with $\Gn$ denoting the Newtonian constant of gravitation, $\ell$ being a constant such that $[\ell] = \mbox{length}$, $\rho (\bm{x})$ is the mass density of the system, and $1\leq s < 3/2$ denotes the fractional parameter. It is often useful to deal with the fractional Poisson equation in the momentum space, hence taking the Fourier transform of both sides of Eq.~\eqref{eq:fractionalpoisson} yields
\be
\label{eq:fractionalpoisson-fourier}
\wh{\Phi} (\bm{k}) = - \frac{4 \, \pi \, \Gn \, \ell ^{2 - 2 s}}{|\bm{k}|^{2s}} \, \wh{\rho}(\bm{k}) \, .
\ee

\begin{remark}
Note that Eq.~\eqref{eq:fractionalpoisson-fourier} allows one to justify the condition $s<3/2$ on $\Rset ^3$, see \eg, \cite{regularity} for details.
\end{remark}

If one considers the case of a point-like source of mass density $\rho (\bm{x}) = \delta ^{3} (\bm{x})$ then one finds
\be
\label{eq:pot-particle-1}
\Phi _s (\bm{x}) = 
- \frac{\Gamma \left(\frac{3}{2} - s \right)}{4^{s-1} \, \sqrt{\pi} \, \Gamma (s)} 
\left( \frac{\ell}{|\bm{x}|} \right)^{2-2s} \, \frac{\Gn \, M}{|\bm{x}|} \, , \qquad \mbox{for} \,\, 1 \leq s < \frac{3}{2} \, .
\ee
Clearly, this expression is not well-behaved as $s \to (3/2)^-$, as expected. However, focusing on the momentum-space representation of the fractional Poisson Eq.~\eqref{eq:fractionalpoisson-fourier} for $s=3/2$, \ie,
\be
\wh{\Phi} (\bm{k}) = - \frac{4 \, \pi \, \Gn \, M}{\ell \, |\bm{k}|^{3}} \, ,
\ee
the inverse Fourier transform of which can be regularized (see \cite{mine}) and yields
\be
\label{eq:pot-particle-2}
\Phi _{3/2} (\bm{x}) \overset{{\rm reg}}{=} \frac{2 \, \Gn \, M}{\pi \, \ell} \, \log \left( |\bm{x}|/\ell \right) \, .
\ee
From $\bm{a} = - \bm{\nabla} \Phi _s (\bm{x}) $, and after recalling that 
\be
a (r) = \frac{v(r) ^2}{r} = |\bm{\nabla} \Phi_s (r)| \, ,
\ee
with $r=|\bm{x}|$, one finds the expression of the orbital speed of a test particle around the center as a function of $r$ and $s$, \ie,
\be
\label{eq:vsdot}
v_s (r) = 
\left\{
\begin{aligned}
& \frac{2^{\frac{3}{2} - s}}{\sqrt[4]{\pi}} \sqrt{\frac{\Gamma \left(\frac{5}{2} - s \right)}{\Gamma (s)}}
\left( \frac{\ell}{r} \right)^{1-s} \sqrt{\frac{\Gn M}{r}} \, , \,\,\, \mbox{for} \,\,  1 \leq s <3/2 \, ,\\
& \\
& \sqrt{\frac{2 \, \Gn \, M}{\pi \, \ell}} \, , \,\, \mbox{for} \,\, s = 3/2 \, .
\end{aligned}
\right.
\ee 
This suggests that, in order to smoothly reproduce the flattening of Galaxy rotation curves in fractional Newtonian gravity one needs to turn the theory into a variable-order. This is achieved by replacing $s$ with a scale-dependent fractional parameter $s (r/\ell)$ such that $s (r/\ell) \to 1$ for $r < \ell$ whereas $s (r/\ell) \to (3/2)^-$ as $r \gtrsim \ell$. 

	Note that, differently from pure MOND, this approach is equipped with a critical length scale $\ell$ rather than an acceleration scale $a_0$. Furthermore, even a variable-order version of Eq.~\eqref{eq:fractionalpoisson} yields a linear theory, whereas MOND is inherently non-linear in nature \cite{Bekenstein:1984tv}. However, one can reconcile the phenomenology of the two theories at Galactic scales by means of the (empirical) {\em Tully--Fisher relation} \cite{TF}
$$
v^4 = \Gn \, M \, a_0 \, ,
$$
which leads to
\be
\label{eq:ell}
\ell = \frac{2}{\pi} \sqrt{\frac{\Gn \, M}{a_0}} \, ,
\ee
with $a_0$ denoting the critical acceleration scale of MOND.

Note that in \cite{mine} the scheme of fractional Newtonian gravity and the corresponding MOND--like scenario have been generally put in connection with bootstrapped-Newtonian and corpuscular gravity (see, {\em e.g.}, \cite{corp-1, corp-2, corp-3, corp-4, corp-5, corp-6, corp-7} for details and \cite{corp-review} for a review on the topic).

%
%
%
%

\section{The Kuzmin disk in Fractional Newtonian Gravity}
\label{sec:Kuzmin}

\begin{figure}[h!]
    \centering
    \begin{minipage}{0.45\textwidth}
        \centering
        \includegraphics[width=0.9\textwidth]{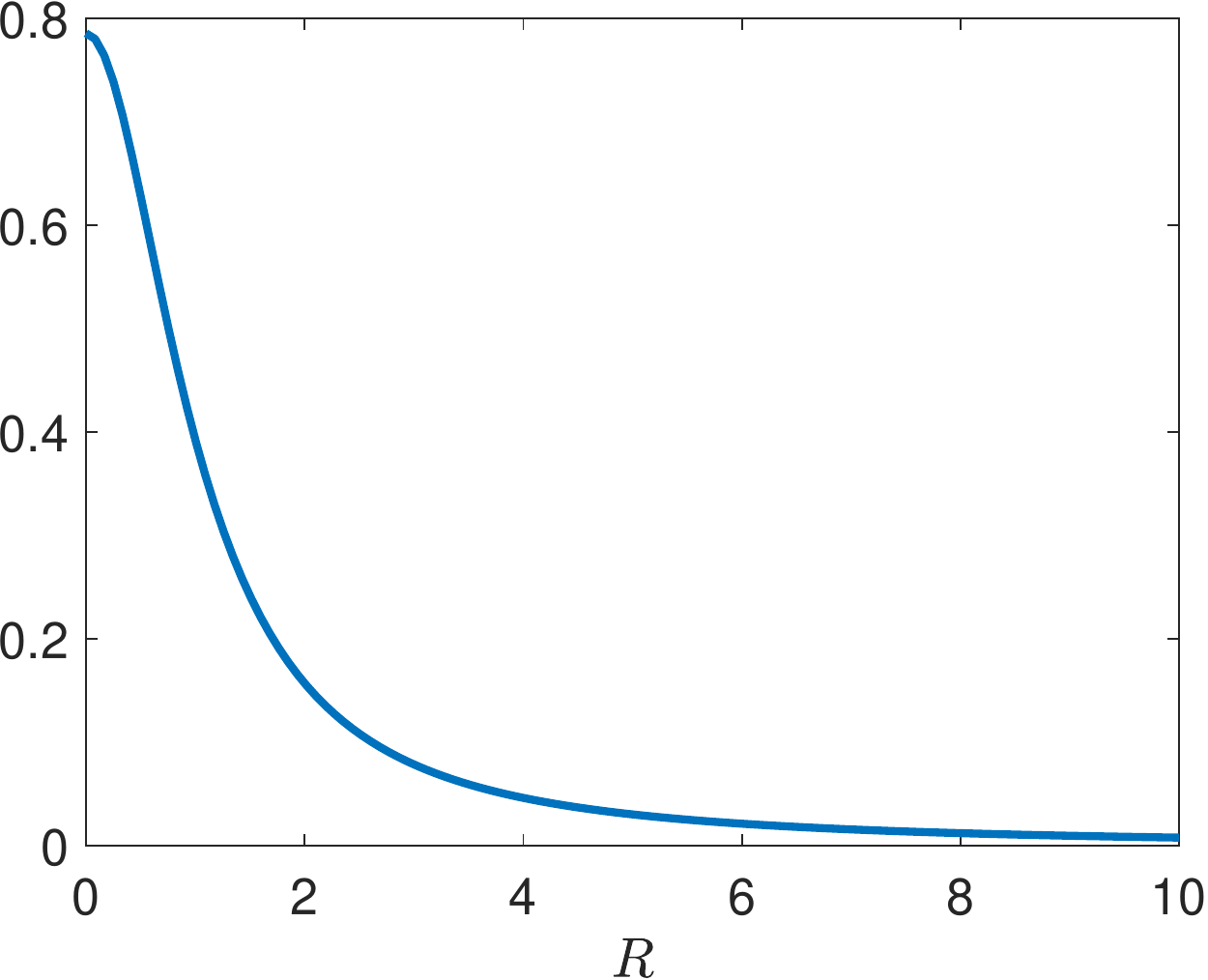}
    \end{minipage}
    \begin{minipage}{0.45\textwidth}
        \centering
        \includegraphics[width=0.9\textwidth]{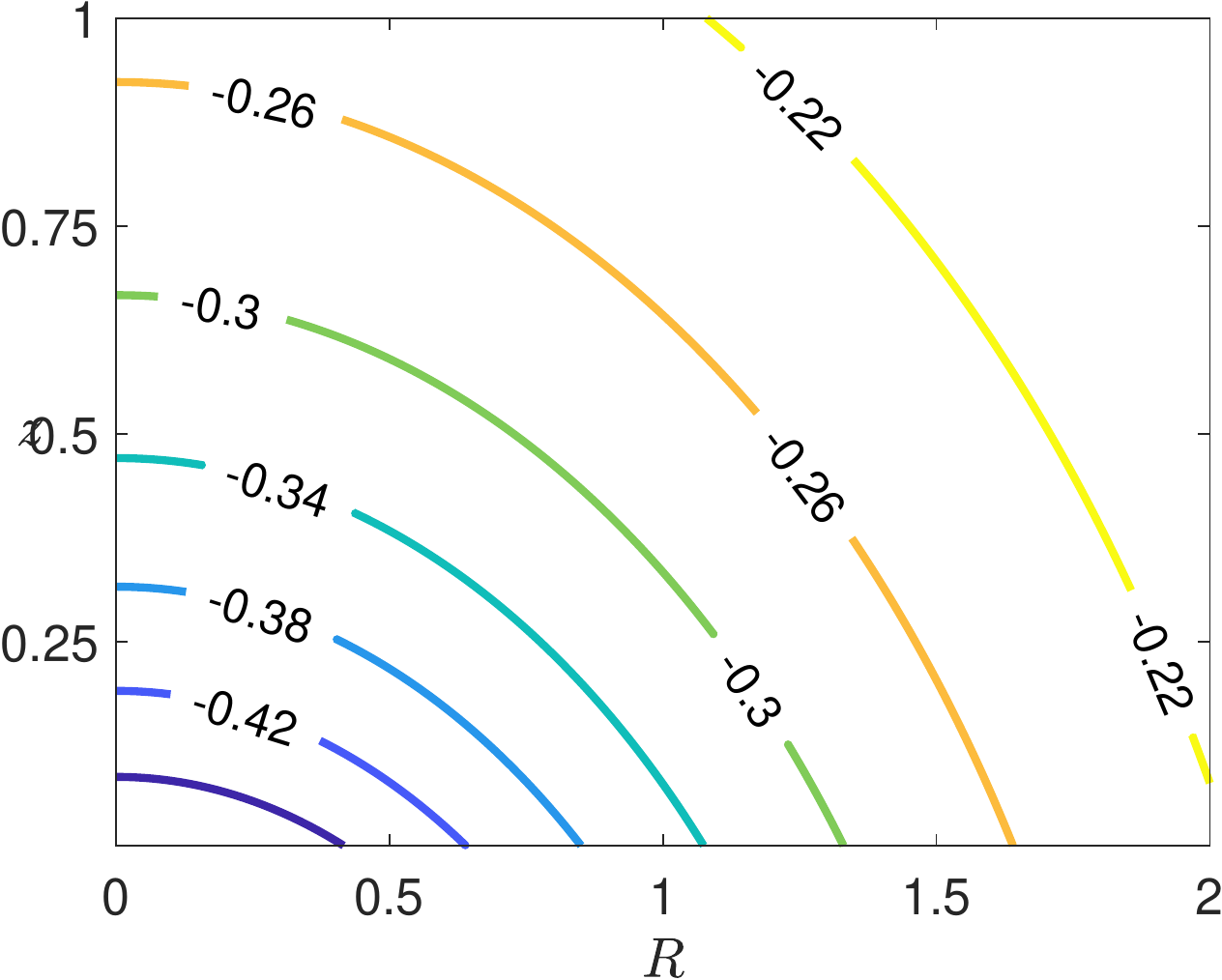}
    \end{minipage}
\caption{Left: mass density of the Kuzmin disk on the plane of the disk ($z=0$) as a function of $R$. Right: Section of the three-dimensional equipotential surfaces ($\Phi _{\rm N} (R, z) = \mbox{const.}$) for the Kuzmin disk. Assumptions: $\Gn = 1$, 
$M = 0.5$, $R_0 = 1$. \label{fig:fig1}}
\end{figure}

The Kuzmin mass density, that in cylindrical coordinates reads
\be
\label{kuzmin-density}
\rho (R, z) = \frac{R_0 \, M}{2 \pi \, (R^2 + R_0^2)^{3/2}} \, \delta (z) \, , 
\ee
with $R_0>0$ and $[R_0] = {\rm length}$, is a widely used axisymmetric model for thin disk Galaxies \cite{BT}. The classical Newtonian solution of Eq.~\eqref{eq:fractionalpoisson} for the Kuzmin disk corresponds to the case $s=1$ and yields
\be
\label{eq:classicalKuzmin}
\Phi _{\rm N} (R, z) \equiv \Phi _{s=1} (R, z) = - \frac{\Gn \, M}{\sqrt{R^2 + (R_0 + |z|)^2}} \, ,
\ee 
see Figure~\ref{fig:fig1}.

Since the Fourier transform of the Kuzmin density reads
$$
\wh{\rho} (\bm{k}) = \wh{\rho} (\kappa) = M \, \eu ^{- \kappa  R_0} \, ,
$$
with $\kappa = \sqrt{k_x^2 + k_y^2}$,
then Eq.~\eqref{eq:fractionalpoisson} for the Kuzmin disk, in the momentum-space, reduces to
\be
\wh{\Phi} (\bm{k}) = \wh{\Phi} (\kappa, k_z)  
= - 4 \, \pi \, \Gn \, \ell ^{2 - 2 s} \, \frac{\wh{\rho} (\bm{k})}{|\bm{k}|^{2s}} 
= - 4 \, \pi \, \Gn \, M \, \ell ^{2 - 2 s}\, \frac{\eu^{- \kappa R_0}}{(\kappa^2 + k_z ^2)^s} \, .
\ee
Inverting $\wh{\Phi} (\bm{k})$ back to position-space yields
\be
\label{eq:KuzminIntegral-1}
\Phi _s (R,z) = - \frac{\Gn \, M \, \ell ^{2 - 2 s}}{\pi}
\int _0 ^\infty \d \kappa \, \kappa \, \eu^{-\kappa R_0} \, J_0 (\kappa R)
\int_{\Rset} \d k_z \,  
\frac{\eu^{i k_z z}}{(\kappa^2 + k_z ^2)^s} \, .
\ee

In order to derive an expression for $\Phi _s (R,z)$ that is more easily treatable from a numerical perspective, one first has to consider the integral
\be
\label{eq:I1}
I_1 (s \, ; \, \kappa, z) = \int_{\Rset} \d k_z \,  
\frac{\eu^{i k_z z}}{(\kappa^2 + k_z ^2)^s} \, .
\ee
Taking advantage of Euler's formula $\eu^{i x} = \cos (x) + i \sin(x)$ and of the known symmetry properties of trigonometric functions one can easily conclude that
\be
I_1 (s \, ; \, \kappa, z) = 2 \int_0^\infty  
\frac{\cos(k_z z)}{(\kappa^2 + k_z ^2)^s} \, \d k_z  \, .
\ee
Then recalling that \cite{abram}
\be
K _{\nu} (x z) = \frac{\Gamma \left(\nu + \frac{1}{2}\right) (2 z)^\nu}{\sqrt{\pi} x^\nu} \, 
\int _0 ^\infty \frac{\cos (x t)}{(t^2 + z^2)^{\nu + \frac{1}{2}}} \d t \, ,
\ee
with $K _{\nu} (z)$ denoting the modified Bessel function of the second kind, $\Re (\nu) > -1/2$, $x >0$, 
and ${\rm arg} (z) < \pi / 2$, one infers that
\be
I_1 (s \, ; \, \kappa, z) = \frac{2^{\frac{3}{2} - s} \sqrt{\pi}}{\Gamma (s)} \left( \frac{|z|}{\kappa} \right)^{s-\frac{1}{2}}
K_{s - \frac{1}{2}} (\kappa |z|) \, .
\ee
Hence, for $z\neq0$ one can rewrite Eq.~\eqref{eq:KuzminIntegral-1} as
\be
\label{eq:KuzminIntegral-2}
\quad
\nonumber
\Phi _s (R,z) = 
- \frac{2^{\frac{3}{2} - s} \, \Gn \, M \, \ell ^{2 - 2 s} |z|^{s-\frac{1}{2}}}{\sqrt{\pi} \, \Gamma (s)} \,
 I_2 (s \, ; \, R, z)
 \, ,
\ee
with
\be
\label{eq:I2}
I_2 (s \, ; \, R, z) = 
\int _0 ^\infty \d \kappa \, 
\kappa^{\frac{3}{2} - s} \, \eu^{-\kappa R_0} \, J_0 (\kappa R) \, K_{s - \frac{1}{2}} (\kappa |z|) \, .
\ee

\begin{remark}
The case $z=0$ has already been analyzed in \cite{mine} and yields
\be
\label{kuzmin-potential-sno32-z0}
\Phi _s (R,0) &=&
\nonumber - \frac{\Gn \, M \, \ell ^{2 - 2 s}}{\sqrt{\pi} \, R_0 ^{3-2 s}} \, 
\frac{\Gamma (s - 1/2) \, \Gamma (3-2 s)}{\Gamma (s)}\\
& \, &  \times \, 
{}_2 F_1 \left(\frac{3}{2}-s, \, 2 - s; \, 1 \, ; \, - \frac{R^2}{R_0^2} \right) \, ,
\ee
with 
$
{}_2 F_1(a, \, b; \, c \, ; \,z)
$
the Gaussian hypergeometric function (see \cite{abram} for details) and $1\leq s < 3/2$. Moreover, the {\em regularized} potential on the plane of the disk for $s=3/2$ reads
\be
\label{kuzmin-potential-s32-z0}
\Phi _{3/2} (R,0) &\overset{\rm reg}{=}& \nonumber
\frac{2}{\pi} \frac{\Gn \, M}{\ell} \,
\log \left[ 1 + \sqrt{1+ \left( \frac{R}{R_0} \right)^2} \,  \right] \, .  
\ee
\end{remark}

\begin{figure}[h!]
   \centering
	\begin{tabular}{cc}
         \includegraphics[width=0.40\textwidth]{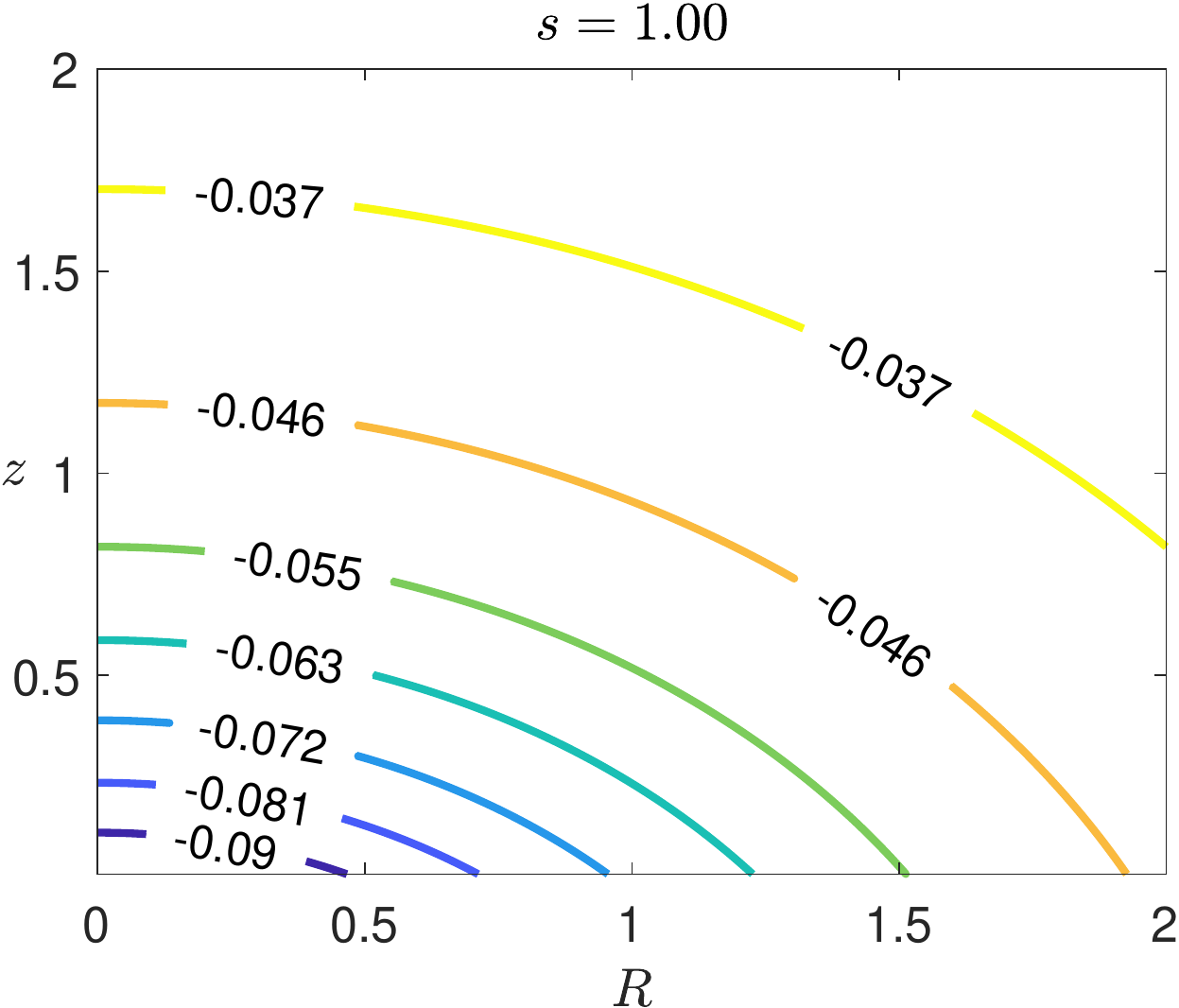} &
         \includegraphics[width=0.40\textwidth]{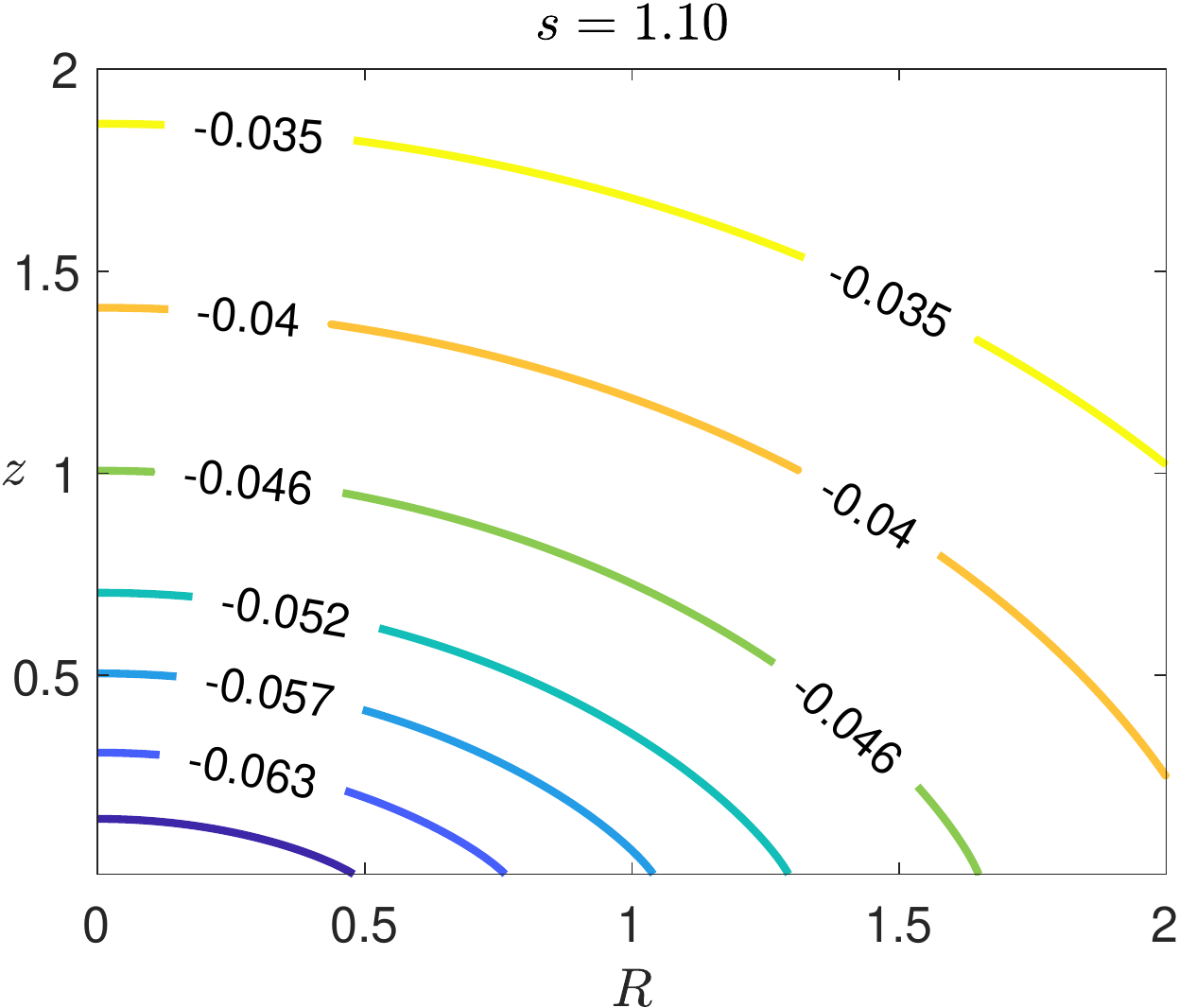}	\\
         \includegraphics[width=0.40\textwidth]{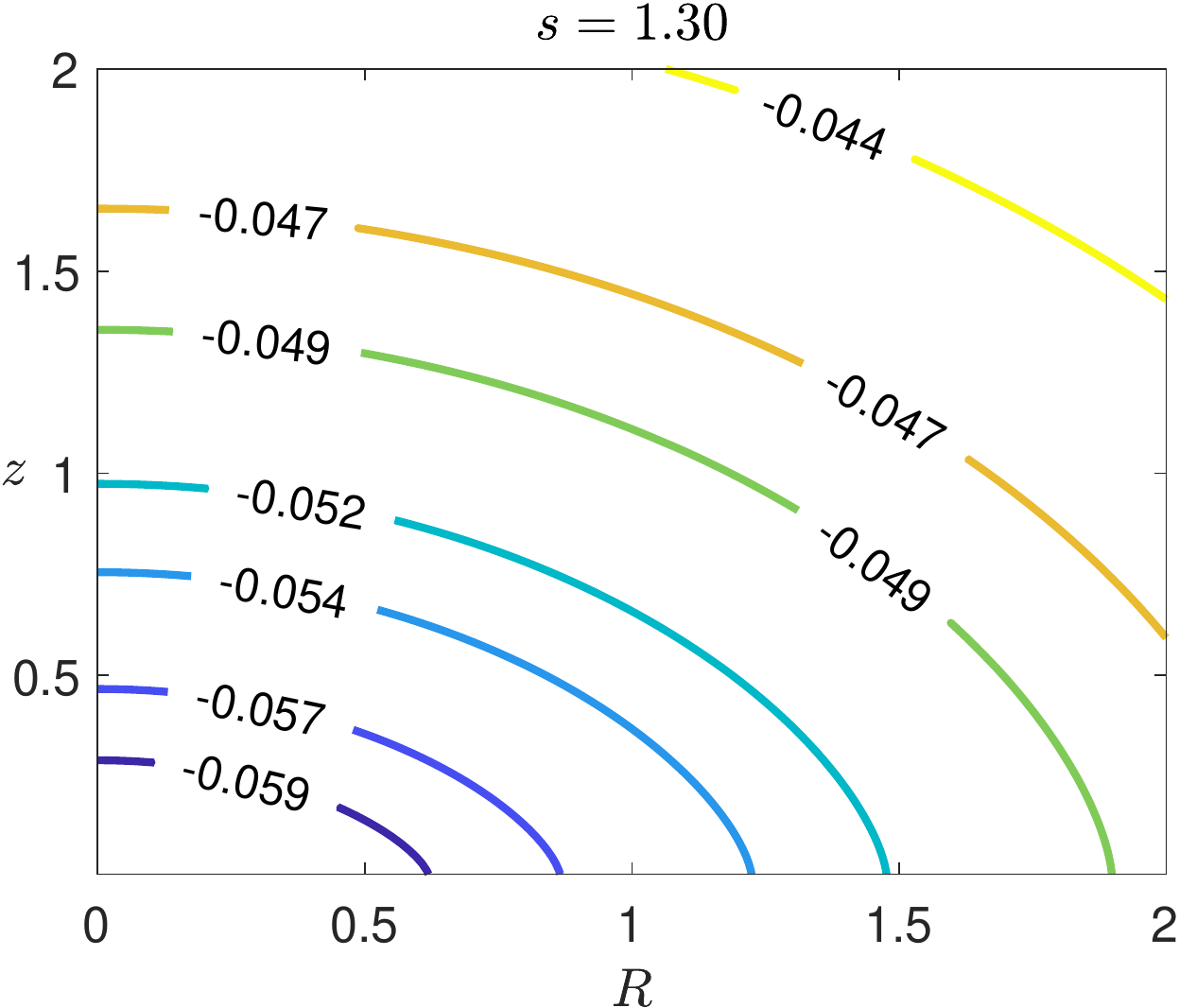} &
         \includegraphics[width=0.40\textwidth]{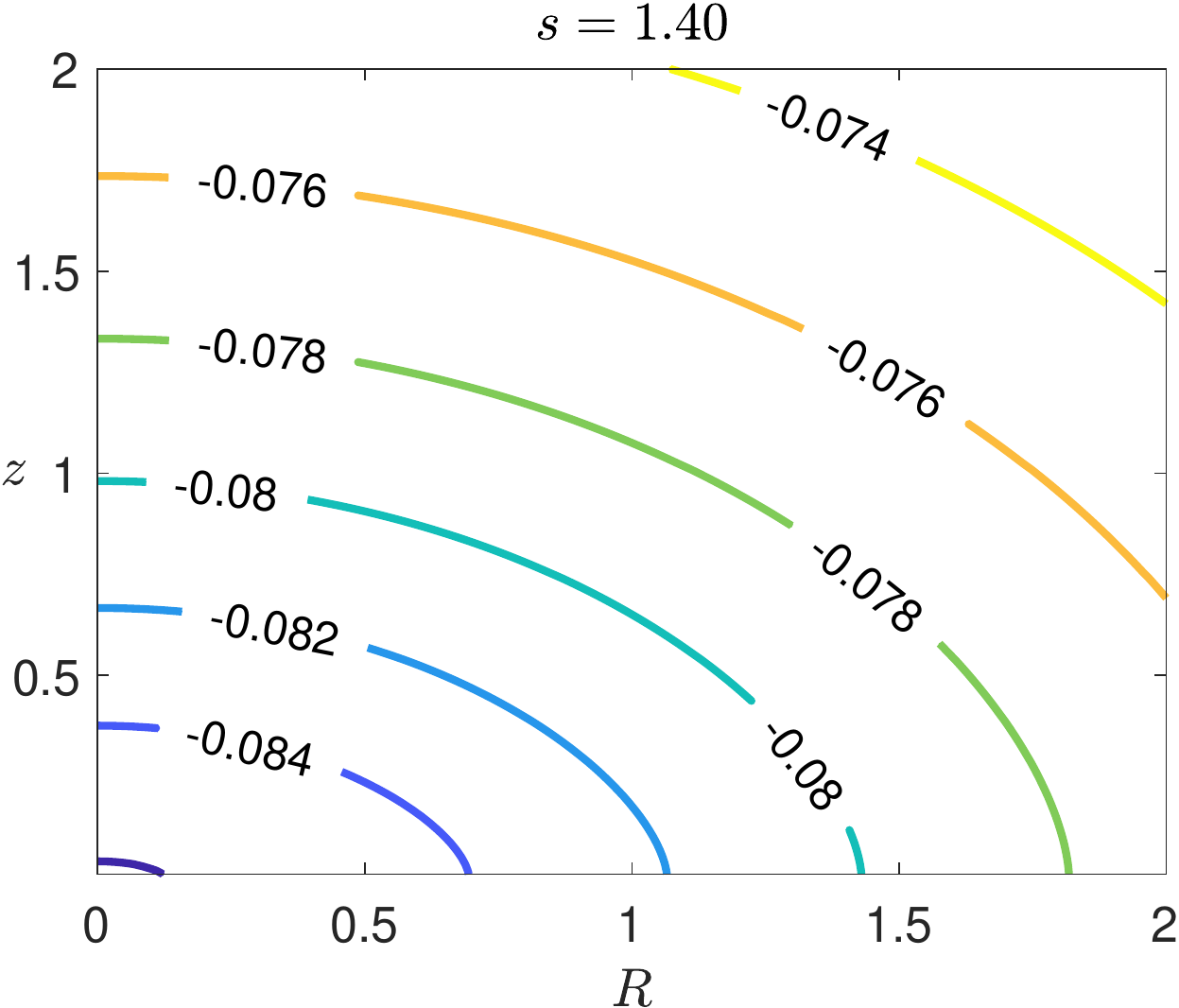}	\\
	\multicolumn{2}{c}{\includegraphics[width=0.40\textwidth]{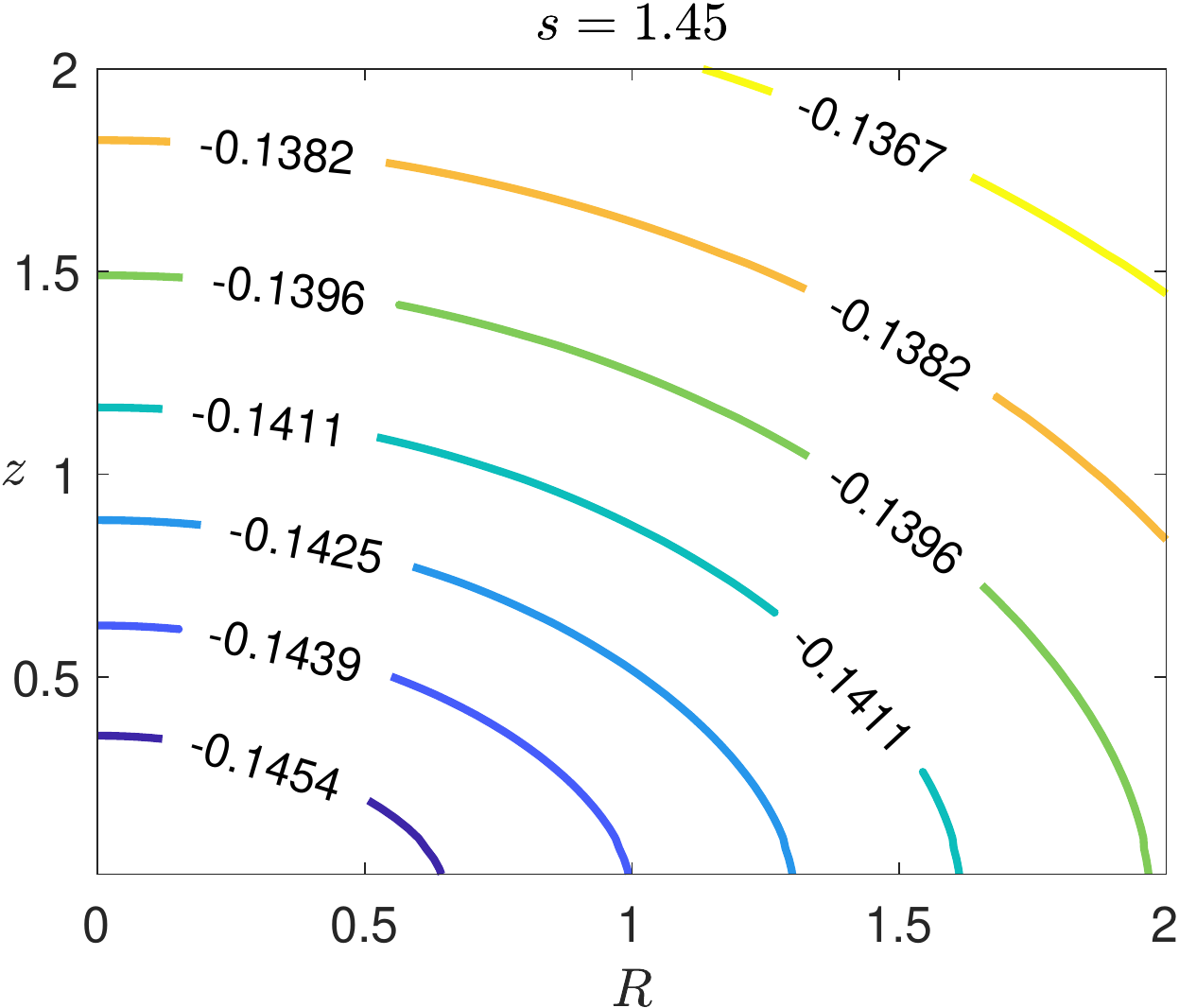}} \
	\end{tabular}
\caption{Sections of $\Phi _s (R, z) = \mbox{const.}$ for positive $R$  
and $z$,
assuming $\Gn \, M = 0.1$, $R_0 = 1$, $\ell = 5$.\label{fig:2}}
\end{figure}

\begin{figure}[h!]
\centering
\includegraphics[width=0.55\textwidth]{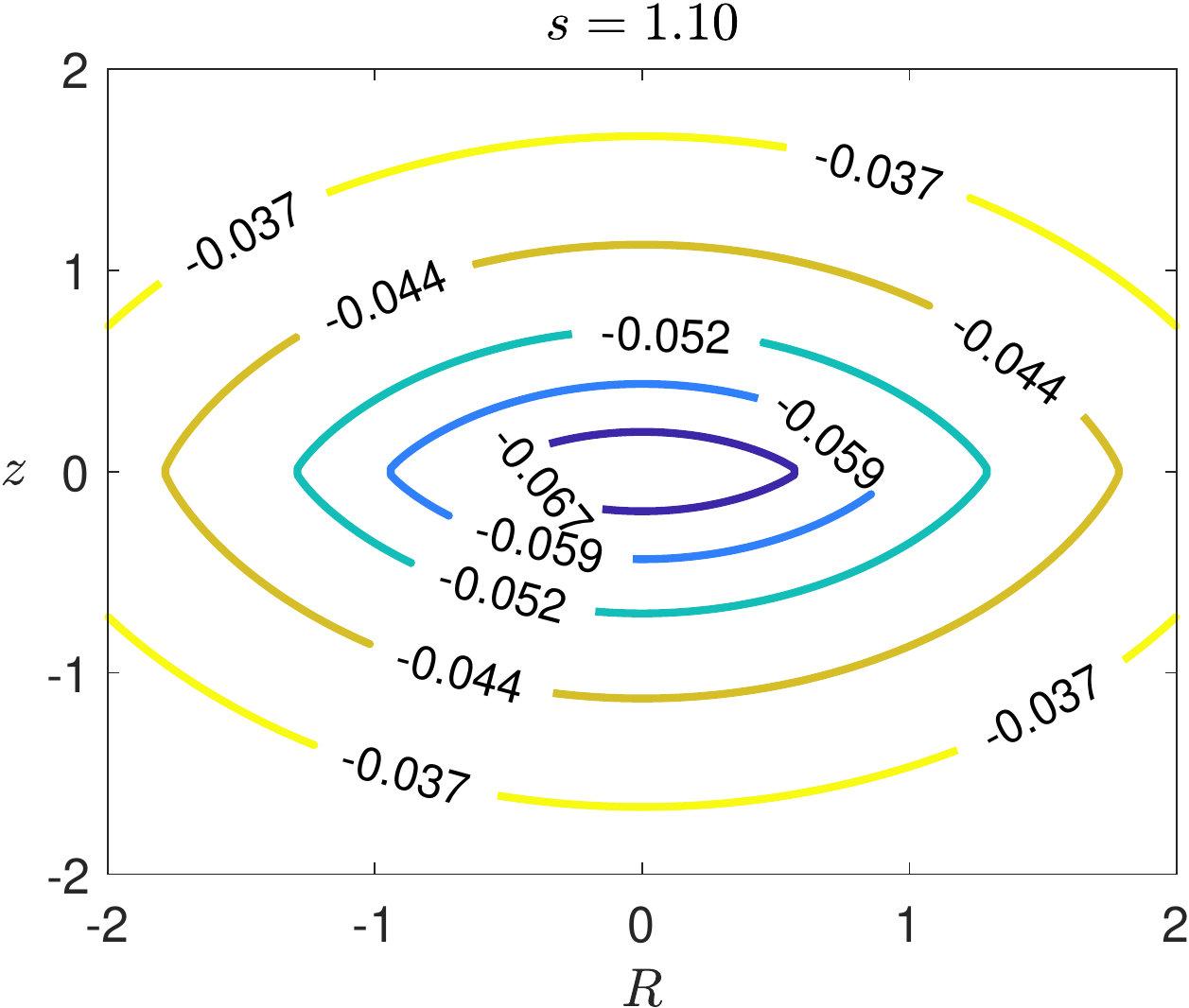}
\caption{Cross section of an equipotential surface with $s=1.1$, assuming $\Gn \, M = 0.1$, $R_0 = 1$, $\ell = 5$.\label{fig:3}}
\end{figure}

The behavior of the potential $\Phi _s (R, z)$ in Eq.~\eqref{eq:KuzminIntegral-2} can be more easily understood thorough the plot of the corresponding equipotential surfaces. Thus, a numerical evaluation of the integral in Eq.~\eqref{eq:KuzminIntegral-2}, based on an adaptive Gauss--Kronrod quadrature, yields the illuminating plots and contours reported in Figures~\ref{fig:2} and~\ref{fig:3}.

%
%
%
%
\subsection{Asymptotic behavior} 
\label{sec:asym}
In \cite{mine} it was shown that for a point particle the solution of Eq.~\eqref{eq:fractionalpoisson} is given by Eq.~\eqref{eq:pot-particle-1} and Eq.~\eqref{eq:pot-particle-2}, where the latter corresponds to $s=3/2$ and it is understood in the regularized sense. Thus, moving away from the Galaxy center one would expect to find a similar behavior from Eq.~\eqref{eq:KuzminIntegral-2} for $r:= \sqrt{R^2 + z^2} \gg R_0$.

From \cite{abram} one recalls that
\be
J_0 (x) \sim \sqrt{\frac{2}{\pi x}} \, \cos \left( x - \frac{\pi}{4} \right) \, , \quad 
K_{\nu} (x) \sim \sqrt{\frac{\pi}{2 x}} \, \eu^{-x} \, ,
\ee
to the lowest order, when $x \to \infty$. This suggests that when $R, z \gg R_0$
\be
\nonumber
I_2 (s \, ; \, R, z) &\sim& \frac{1}{\sqrt{R |z|}} 
\int _0 ^\infty 
\kappa^{\frac{1}{2} - s} \, \eu^{-\kappa (R_0 + |z|)} \, \cos (\kappa R) \, \d \kappa\\
&\sim&
\frac{\Gamma \left(\frac{3}{2} - s\right)}{\sqrt{R |z|} \, \left[ R^2 + (R_0 + |z|)^2 \right]^{\frac{3 -2s}{4}}} \, 
\cos \left[\frac{3 - 2 s}{2} \, \arctan \left( \frac{R}{R_0 + |z|} \right) \right] \, ,
\ee
that yields $I_2 \sim r^{s-\frac{5}{2}}$ when $R, z \gg R_0$, {\em i.e.,} $r= \sqrt{R^2 + z^2} \gg R_0$. From Eq.~\eqref{eq:KuzminIntegral-2} one concludes that
\be
\Phi _s (R,z) \simeq |z|^{s-\frac{1}{2}} \, I_2 (s \, ; \, R, z) \sim |z|^{s-\frac{1}{2}} \, r^{s-\frac{5}{2}} \sim r^{2 s - 3} \, ,
\ee
assuming for simplicity $\mathcal{O} (R) = \mathcal{O} (|z|)$ as $r \to \infty$, which coincides with the asymptotic behavior of the potential for the point particle Eq.~\eqref{eq:pot-particle-1} for $1 \leq s < 3/2$. Furthermore, it is easy to show using the same procedure discussed above to the {\em Hadamard partie finie} of \eqref{eq:KuzminIntegral-2} for $s=3/2$ (see {\em e.g.}, \cite{Riesz, Estrada}) that $\Phi _{3/2} (R,z) \sim \log (r)$ as $r \to \infty$ with $\mathcal{O} (R) = \mathcal{O} (|z|)$.

%
%
%
%
\subsection{Full potential outside the Galactic plane: a series representation}

From \cite{abram} one recalls that
\be
J_0 (x) = \sum _{n=0} ^\infty \frac{(-1)^n}{(n!)^2} \left( \frac{x}{2} \right)^{2n} \, .
\ee
Taking advantage of Lebesgue's dominated convergence theorem one can 
expand $J_0$ in Eq.~\eqref{eq:I2} and interchange the summation and integral. This leads to
\be
\label{eq:I2-sum}
\nonumber
I_2 (s \, ; \, R, z) &=& 
\sum _{n=0} ^\infty \frac{(-1)^n}{(n!)^2} \left( \frac{R}{2} \right)^{2n}
\int _0 ^\infty 
\kappa^{\frac{3}{2} + 2n - s} \, \eu^{-\kappa R_0} \, K_{s - \frac{1}{2}} (\kappa |z|) \, \d \kappa \\
&\equiv& \sum _{n=0} ^\infty \frac{(-1)^n}{(n!)^2} \left( \frac{R}{2} \right)^{2n} \, I_3 (s , n \, ; \, R, z) \, ,
\ee
with
\be
\label{eq:I3}
I_3 (s , n \, ; \, R, z) := \int _0 ^\infty 
\kappa^{\frac{3}{2} + 2n - s} \, \eu^{-\kappa R_0} \, K_{s - \frac{1}{2}} (\kappa |z|) \, \d \kappa \, .
\ee
If one recalls the definitions of Kummer's (confluent hypergeometric) functions \cite{abram}
\be
M (a,b,z) = \sum_{n=0}^\infty \frac{(a)_n}{(b)_n} \frac{z^n}{n!} \equiv {{}_1 F}_1 (a;b;z) \, ,
\ee
and
\be
U(a,b,z) = \frac{\Gamma (1-b)}{\Gamma (a-b+1)} \, M (a,b,z) + 
\frac{\Gamma (b-1)}{\Gamma (a)} \, z^{1-b} \, 
M (a-b+1,2- b,z) \, ,
\ee
it is not hard to see that 
\be
K_{\nu} (z) = \sqrt{\pi} (2 z)^\nu \, \eu^{-z} \, U\left(\nu +\frac{1}{2}, 2 \nu + 1, 2z\right) \, .
\ee
The last expression for the modified Bessel function of the second kind then implies that
\be
K_{s - \frac{1}{2}} (\kappa |z|) = \sqrt{\pi} \, (2 \kappa |z|)^{s - \frac{1}{2}} \, \eu^{- \kappa |z|} \, 
U\left( s, 2s, \, 2 \kappa |z| \right) \, ,
\ee
that once inserted in Eq.~\eqref{eq:I3} allows one to rewrite $I_3$ as
\be
\label{eq:I3-Kummer}
I_3 (s , n \, ; \, R, z) = \sqrt{\pi} \, (2 |z|)^{s - \frac{1}{2}}
\int _0 ^\infty 
\kappa^{2n + 1} \, \eu^{-\kappa ( R_0 + |z|)} \, U\left( s, 2s, \, 2 \kappa |z| \right) \, \d \kappa \, .
\ee
If one recalls the known special integral (see, {\em e.g.} \cite[\S 13.10(ii), Eq. 13.10.7]{NIST})
\be
\qquad \int _0 ^\infty \eu^{- z t} \, t^{b-1} \, U(a,c, \, t) \, \d t = 
\frac{\Gamma (b) \, \Gamma (b - c + 1)}{\Gamma (a + b -c +1) \, z^{b}} \, 
{{}_2 F}_1 \left( a, b; \, a+b-c+1 \, ; \, \frac{z - 1}{z} \right) \, ,
\ee
with $\Re (b) > \max \{ \Re (c) - 1 \, , \,  0 \}$ and $\Re (z) > 0$, then Eq.~\eqref{eq:I3-Kummer} reduces to
\be
\label{eq:I3-result}
\nonumber
I_3 (s , n \, ; \, R, z) &=& \frac{\sqrt{\pi} (2 |z|)^{s - \frac{1}{2}}}{(R_0 + |z|)^{2n + 2}} \frac{\Gamma (2n + 2) \Gamma (2n - 2s + 3)}{\Gamma (2n + 3 - s)} \times\\
&& \quad \times {{}_2 F}_1 \left( s, 2n+2; 2n + 3 - s \, ; \, \frac{R_0 - |z|}{R_0 + |z|} \right) \, .
\ee

Therefore, combining Eq.s~\eqref{eq:KuzminIntegral-2}, \eqref{eq:I2-sum}, and \eqref{eq:I3-result} one finds
\be
\nonumber
\Phi _s (R,z) &=& 
- \frac{2^{\frac{3}{2} - s} \, \Gn \, M \, \ell ^{2 - 2 s} |z|^{s-\frac{1}{2}}}{\sqrt{\pi} \, \Gamma (s)} \,
 I_2 (s \, ; \, R, z) \\
\nonumber &=&
- \frac{2 \, \Gn \, M \, \ell ^{2 - 2 s} |z|^{2 s-1}}{\Gamma (s) \, (R_0 + |z|)^2} \times \\
\nonumber & & \quad \times \sum _{n=0} ^\infty \frac{(-1)^n}{(n!)^2} \left[ \frac{R}{2 (R_0 + |z|)} \right]^{2n}
\frac{\Gamma (2n + 2) \Gamma (2n - 2s + 3)}{\Gamma (2n + 3 - s)} \times \\
&& \qquad \qquad \times {{}_2 F}_1 \left( s, 2n+2; 2n + 3 - s; \, \frac{R_0 - |z|}{R_0 + |z|} \right)
 \, ,
\label{eq:schifo}
\ee
that ultimately provides an explicit expression of the potential, outside of the Galactic plane, in terms of a series of known special functions. However, such an expression can hardly be useful when dealing with observations, thus the numerical evaluation of Eq.s~\eqref{eq:KuzminIntegral-2} and \eqref{eq:I2} turns out to be a more practical path to follow.
 
%
%
%
%

\section{Conclusions and outlook}

	Fractional Newtonian gravity \cite{mine}, based on the fractional extension of Poisson's field equation for the gravitational potential obtained trough the replacement of the Laplacian with the fractional Laplacian, represents a novel application of fractional calculus to astrophysics. Most notably, this theory naturally comprise both Newtonian gravity and MOND's asymptotic behavior as limiting scenario, respectively obtained setting $s=1$ and $s=3/2$. This particular feature surprisingly allows one to naturally connect observations of Galaxy rotation curves with the more abstract theory of weakly-singular integro-differential operators, and hence to non-local theories of gravity.

	In this work we have completed the analysis for an important toy model for the mass distribution of very thin-disk galaxies, known as the Kuzmin disk. First, in Eq.~\eqref{eq:KuzminIntegral-2} we have provided an explicit integral representation of the potential generate by the disk outside the plane of the disk. Second, we have computed numerically the form of the equipotential surfaces for different values of the fractional parameter $s$ and we provided some illuminating cross sections of these surfaces in Figure~\ref{fig:2} and~\ref{fig:3}. Third, in Section~\ref{sec:asym} we verified the asymptotic behavior of the potential in Eq.~\eqref{eq:KuzminIntegral-2} when $r \to \infty$. Finally, in Eq.~\eqref{eq:schifo} we have provided an explicit series representation for the potential generated by the Kuzmin disk, outside the plane of the disk $z=0$, thus filling a gap in the literature.

	The program of fractional Newtonian gravity surely looks promising and deserving of further investigation. First, in order to properly reproduce Galaxy rotation curves one needs to turn the theory into a {\em variable-order one}, with $s=s(\bm{x}/\ell)$ being a function reducing to $1$ at short-scale and approaching $3/2$ as one moves asymptotically far away from the center of the Galaxy. However, promoting this model to a variable-order theory lead to complications (see, {\em e.g.}, \cite{samkoNODY}), both mathematical and numerical. These more serious topics will be discussed in detail in future studies.  

%
%
%
%

\section*{Acknowledgments}	

The work of Roberto Garrappa is supported by a GNCS-INdAM 2020 Project. Andrea Giusti and Genevi{\`e}ve Vachon are supported by the Natural Sciences and Engineering Research Council of Canada (Grant No.~2016-03803 to V. Faraoni) 
and by Bishop's University. The work of Andrea Giusti  has been carried out in the framework of the activities of the Italian National Group for Mathematical Physics [Gruppo Nazionale per la Fisica Matematica (GNFM), Istituto Nazionale di Alta
Matematica (INdAM)].

\vskip 1 cm

\end{document}